\documentclass[sigconf]{acmart}

\AtBeginDocument{%
  \providecommand\BibTeX{{%
    \normalfont B\kern-0.5em{\scshape i\kern-0.25em b}\kern-0.8em\TeX}}}

\setcopyright{acmcopyright}
\copyrightyear{2020}
\acmYear{2020}

\acmConference[]{}{}{}
\acmBooktitle{}
\acmPrice{15. 00}
\acmISBN{978-1-4503-9999-9/18/06}

\usepackage{array}
\newcolumntype{M}[1]{>{\centering\arraybackslash}m{#1}}

\usepackage{caption}
\usepackage{subcaption}
\usepackage{tablefootnote}
\usepackage{algorithm}
\usepackage{dblfloatfix}    
\usepackage[noend]{algpseudocode}
\hypersetup{draft}

\begin{document}

\title[Hashtags are (not) judgemental]{
   Hashtags are (not) judgemental: The untold story of Lok Sabha elections 2019
}







\author{Saurabh Gupta}
\affiliation{\institution{IIIT-Delhi}}
\email{saurabhg@iiitd.ac.in}

\author{Asmit Kumar Singh}
\affiliation{\institution{IIIT-Delhi}}
\email{asmit18025@iiitd.ac.in}

\author{Arun Balaji Buduru}
\affiliation{\institution{IIIT-Delhi}}
\email{arunb@iiitd.ac.in}

\author{Ponnurangam Kumaraguru}
\affiliation{\institution{IIIT-Delhi}}
\email{pk@iiitd.ac.in}

\renewcommand{\shortauthors}{S. Gupta et al.}

\begin{abstract}
    Hashtags in online social media have become a way for users to build communities around topics, promote opinions, and categorize messages. In the political context, hashtags on Twitter are used by users to campaign for their parties, spread news, or to get followers and get a general idea by following a discussion built around a hashtag. In the past, researchers have studied certain types and specific properties of hashtags by utilizing a lot of data collected around hashtags. In this paper, we perform a large-scale empirical analysis of elections using only the hashtags shared on Twitter during the 2019 Lok Sabha elections in India. We study the trends and events unfolded on the ground, the latent topics to uncover representative hashtags and semantic similarity to relate hashtags with the election outcomes. We collect over 24 million hashtags to perform extensive experiments. First, we find the trending hashtags to cross-reference them with the tweets in our dataset to list down notable events. Second, we use Latent Dirichlet Allocation to find topic patterns in the dataset. In the end, we use skip-gram word embedding model to find semantically similar hashtags. We propose popularity and an influence metric to predict election outcomes using just the hashtags. Empirical results show that influence is a good measure to predict the election outcome.
\end{abstract}




\maketitle

\section{Introduction}

Online social media platforms like Twitter are being used by people to spread information and opinions among other users. A lot of times, people are observed reporting the ground events happening near them, making Twitter a source of getting breaking news ~\cite{beatriz, abhinav}. For example, when the terrorist attacks in Mumbai in 2008 were happening, Twitter users in India (especially in Mumbai) were providing an instant eyewitness account of what was happening at the ground \cite{Beaumont}. More recently, a lot of media channels covered the reactions of people all over India using Twitter when article 370 was scraped ~\cite{dhweb, pti}. Twitter is considered so effective that even the Indian government recently asked them to remove accounts spreading rumors about Kashmir ~\cite{rezwan}.

\begin{figure}[!b]
    \centering
    \includegraphics[width=0.8\linewidth]{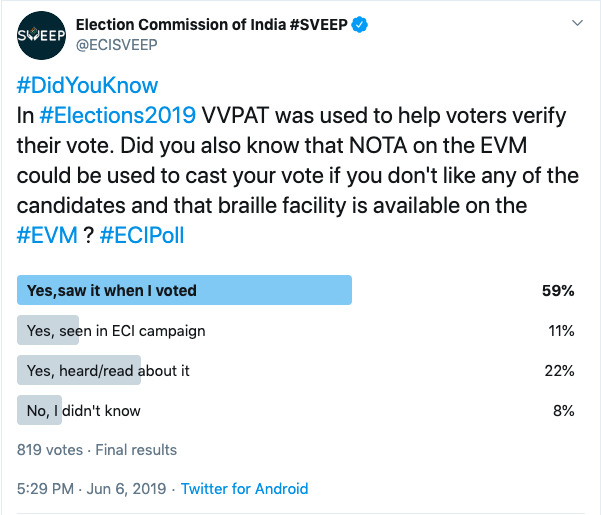}
    \caption{A tweet with multiple hashtags: \#elections2019, \#evm, \#ecipoll. Hashtags aid the formation of ad hoc publics around elections 2019, EVM (Electronic Voting Machine) and ECI (Election Commission of India) poll.}
    \label{fig:sample}
\end{figure}

Hashtags on Twitter are generally used: by many brands to promote a product; by users to categorize their messages, build communities around a topic; and are an efficient way to join public discussions. It is very common to see the use of hashtags to aid the formation of ad-hoc publics around specific events like \#ausvotes, \#londonriots, \#wikileaks\cite{quteprints46515}, and \#elections2019 (see Figure \ref{fig:sample}). 

In the past, hashtags helped users connect to various political movements like the \#iranelection, and the \#blacklivesmatter. More recently, hashtags brought together the users who are concerned about Lok Sabha elections and wanted to share their opinions. Twitter has become one of the most effective unofficial platforms to share news, opinions, facts, fake news, and a political playground with \#LokSabhaElections2019 among the top three most tweeted hashtags in 2019 \cite{nandita}. Hashtags also aided researchers to study such political events from multiple perspectives like participation in \#iranelection ~\cite{Gaffney10iranelection:quantifying}, retweet behavior on real-world ground events ~\cite{ICWSM101484}, temporal and demographic characterstics~\cite{articlemunde}, category and nature of users and tweets ~\cite{Small2011} and so on. 

In this paper, we perform a large-scale empirical study of political hashtags from Indian context on Twitter. 
We start with finding the most trending hashtags over the course of elections and during each phase. We then map these trends to real-world events that happened and were captured in our dataset. Further, we perform topic modeling using Latent Dirichlet Allocation (LDA) on the complete dataset to find out topics during the elections. We also find semantically similar hashtags using the context vectors created using skip-gram word embeddings. Then we search for some key candidates participating in the polls. At last, we use semantically similar hashtags to find out which candidate was more popular on Twitter and compare it with election outcomes (win/lose). During 2019 Lok Sabha elections in India, Twitter was used by a lot of political parties, candidates, party supporters and common people to spread opinions, promotions, campaign, etc. To study the role of Twitter in the elections, we collected data from Feb 05, 2019 to Jun 25, 2019. Our collection process was heavily based on hashtags. We looked at hourly trends in keywords and hashtags to manually filter only the ones that are related to elections. 



We believe our study can help several entities involved in political movements. The trends across all phases during elections can help political parties assess user sentiments towards them over Twitter and help to plan political propaganda. The patterns also facilitate the users to get a slight intuition about what party or candidate is more favored. The events fetched using hashtags give an idea about what is going on around that hashtag. The topics and semantics are majorly dominated by the candidates who heavily use social media to express their opinions. Semantic similarity also reveals the  \#hashtags to which a candidate's or party's name is associated. For example, we (in a completely unsupervised manner) observe that \#modi is getting associated with \#surgicalstrikes. We provide a way to quantify and contrast election outcome with Twitter using only the hashtags. Most of the analysis mentioned above is imaginable when you have a lot of attributes from the tweets. The fact that we only use the bare minimum hashtags for all this makes this study different from others. We present a way to achieve similar results using just the hashtags.


The organization of the rest of the paper is as the following. We first explain our data collection strategy, followed by some initial analysis on hashtags in Section \ref{sec:data}. In Section \ref{sec:trends}, we plot word clouds to show what all hashtags were trending during the elections to analyze a general preference related to elections spread across Twitter. Further, we use the word clouds to get events happening over elections, which might be causing specific hashtags to trend. In Section \ref{sec:tns}, we find topics among the hashtags using LDA and semantically similar hashtags using skip-gram word embeddings. We further use semantically similar hashtags to contrast the battle of two candidates on social media based on the hashtag usage and compare it with election outcomes.
In Section \ref{sec:literature} we mention some of the related work and discuss the findings in Section \ref{sec:conclude} with some observations.

\section{Data Collection and Initial Analysis}
\label{sec:data}
In this section, we first discuss the data collection strategy, then showcase some preliminary analysis on hashtags. 

\subsection{Data Collection}
The Loksabha Elections in India started on Apr 11th, 2019 and ended on May 19th, 2019.\footnote{Except for the Vellore Parliamentary constituency in Tamil Nadu where the Election Commission of India (ECI) canceled the elections ~\cite{manasa}.} We collected tweets from Feb 05, 2019 to Jun 25, 2019 - based on the intuition that people start talking about elections way before the actual dates and go on talking about it several days after it gets over. The elections occurred in seven phases where votes were cast in a single day followed by a few no-voting days. The timeline is shown in Table \ref{tab:duration}.

\begin{table}[htbp]
    \centering
    \begin{tabular}{c|c|c}
    \hline
         \textbf{Phase} & \textbf{Date of voting} & \textbf{Duration of each phase}\tablefootnote{We make the split based on an assumption that when users vote in a phase, they will talk about things related to corresponding phase unless a new phase starts. For example, during phase 1, the trending hashtags will be related to phase 1 unless phase 2 starts. As shown in Figure \ref{fig:top-hash}, the dominant presence of hashtags like \#votinground1, \#votinground2, \#phase5 during these phases support our assumptions.} \\
         \hline
         1 &  Apr 11 & Apr 11 - Apr 17 \\
         2 & Apr 18 & Apr 18 - Apr 23 \\
         3 & Apr 24 & Apr 24 - Apr 28 \\
         4 & Apr 29 & Apr 29 - May 5 \\
         5 & May 6 & May 6 - May 11 \\
         6 & May 12 & May 12 - May 18 \\
         7 & May 19 & May 19 - May 22\tablefootnote{The counting of votes started on May 23. Therefore, we assumed that Phase 7 lasted until May 22.} \\
         \hline
    \end{tabular}
    \caption{Phase-wise election's date and duration.}
    \label{tab:duration}
\end{table}

Initially, we started looking at hourly trends in keywords and hashtags from twenty-two cities in India. We manually selected hashtags related to elections based on the hourly trends and used Twitter's streaming API to get the following posts containing such hashtags. Further, we used Twitter's search API to collect the tweets we missed due to the manual addition. On the date of each phase - Apr 11, Apr 18, and so on,  we performed the same process at an interval of fifteen minutes instead of an hour. We collected a total of 45.1 million tweets, out of which 9.4 million were original tweets, and the rest were retweeted or quoted tweets. On investigation, we found some discrepancies with the collection process. For example, Twitter API failed to parse the hashtags from the text of some tweets. To resolve these discrepancies we: i) parsed the hashtags in instances of tweets where a user inserted a space between the \# and the term. For example, \# elections2019 is parsed as \#elections, and ii) filtered the instances of tweets where the Twitter API failed to capture the hashtags. The parsing and filtering on the 9.4 million tweets resulted in the removal of 1.18 million tweets. In the remaining 8.22 million tweets, there were 24.9 million hashtags. Some statistics about the final 8.22 million tweets are given in Table \ref{tab:stats}. 

\begin{table}[htbp]
    \centering
    \begin{tabular}{c|c}
    \hline
        Total tweets after preprocessing & 8,228,932 \\
        Total hashtags in tweets & 24,958,397 \\
        No. unique hashtags in dataset  & 970,408 \\
        Minimum no. of hashtags in a tweet & 1 \\
        Maximum no. of hashtags in a tweet & 7 \\
        Average no. of hashtags per tweet & 3.02 \\
    \hline
    \end{tabular}
    \caption{Summary Statistics of the dataset. Number of hashtags vary from 1 to 7 with an average of 3.02 hashtags per tweet.}
    \label{tab:stats}
\end{table}

\textbf{Ethical Considerations.} The data we collected is through Twitter's public API. All the data is stored in a central server with restricted access and firewall protection. All the fields from the data are removed except for the timestamp, the text and the hashtags, which are used in this study. All experiments are performed on this dataset. We intend to make the data publicly available upon acceptance.

\subsection{Data Distribution}
\begin{figure}[htb!]
    \centering
    \includegraphics[width=0.8\linewidth]{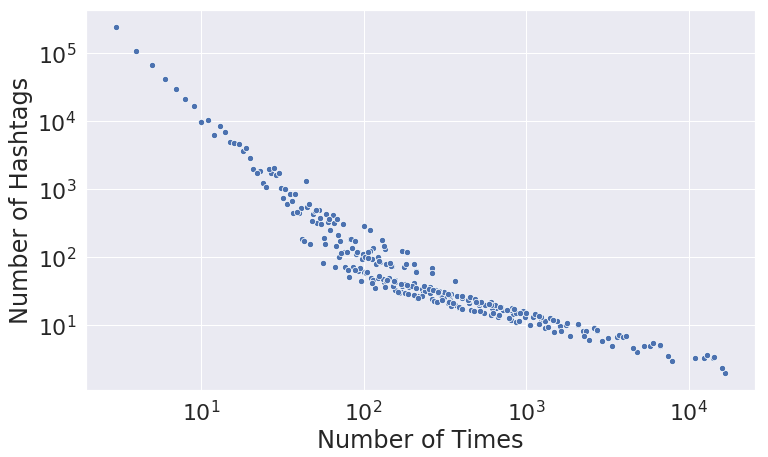}
    \caption{Distribution of number of times a hashtag is shared in the dataset. Both x and y axes are in log scale. There are more number of hashtags that are tweeted lesser number of times, and vice-versa.}
    \label{fig:tvh}
\end{figure}

Figure \ref{fig:tvh} shows the distribution of the number of times hashtags are shared. The distribution follows power law, i.e., most hashtags are shared only a few number of times. 

\begin{figure}[hbt!]
    \centering
    \includegraphics[width=0.8\linewidth]{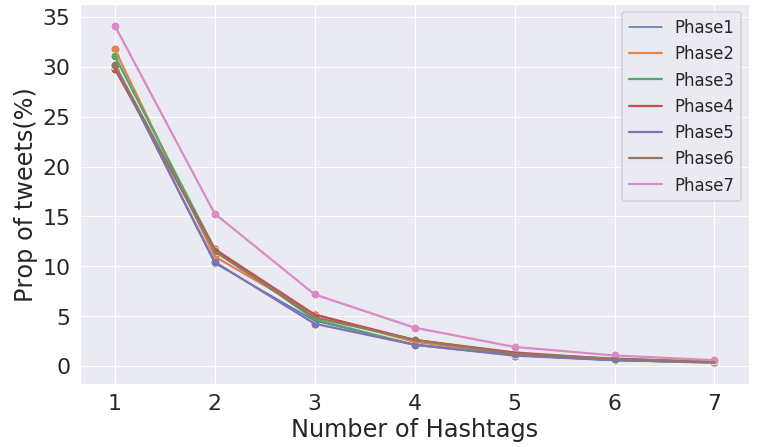}
    \caption{Distribution of number of hashtags in each tweet among all phases. There are ~40-45\% of tweets that contain only one or two hashtags. The number decreases with the increase in the number of hashtags in each tweet.}
    \label{fig:pvh}
\end{figure}

\begin{figure*}[hbtp] 
  \begin{subfigure}[b]{0.30\linewidth}
    \centering
    \includegraphics[width=\linewidth]{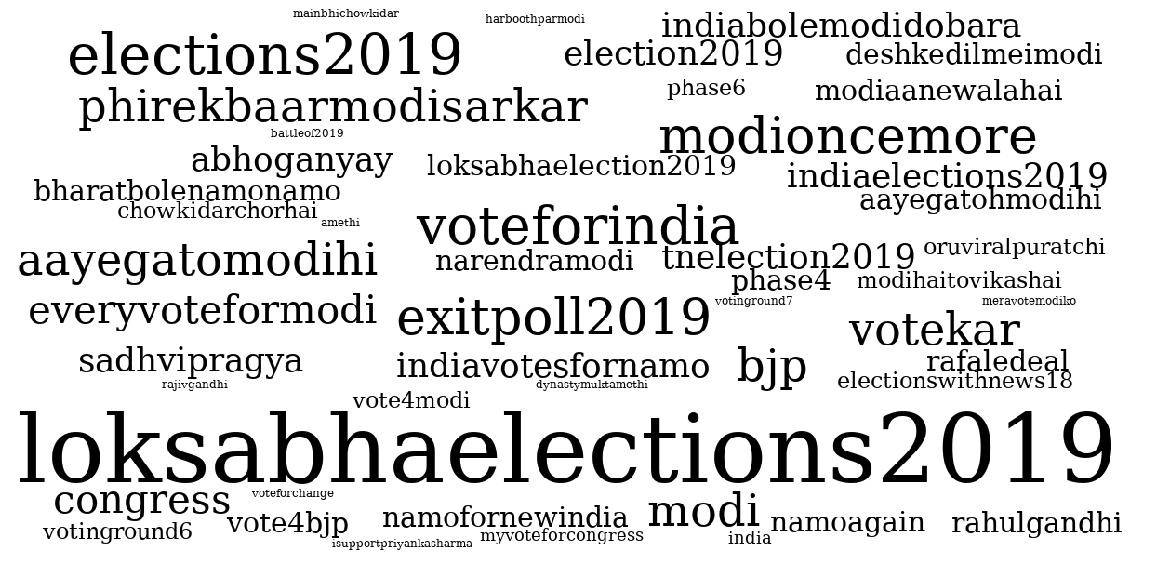}
    \caption{Overall} 
    \label{fig:top-hash-all} 
  \end{subfigure}
  \begin{subfigure}[b]{0.30\linewidth}
    \centering
    \includegraphics[width=\linewidth]{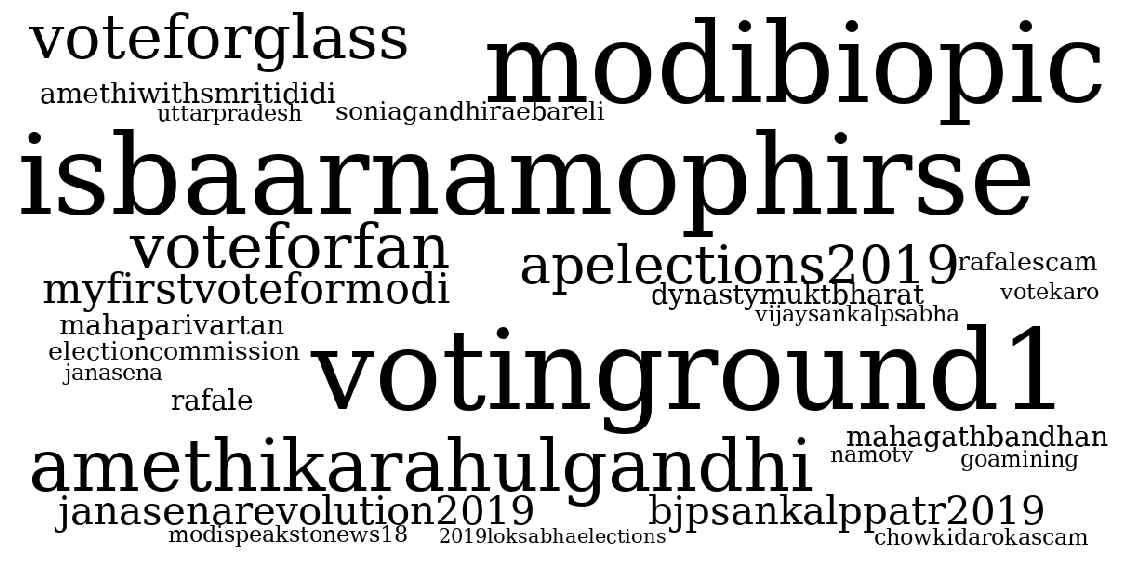}
    \caption{Phase 1} 
    \label{fig:top-hash-1} 
  \end{subfigure} 
  \begin{subfigure}[b]{0.30\linewidth}
    \centering
    \includegraphics[width=\linewidth]{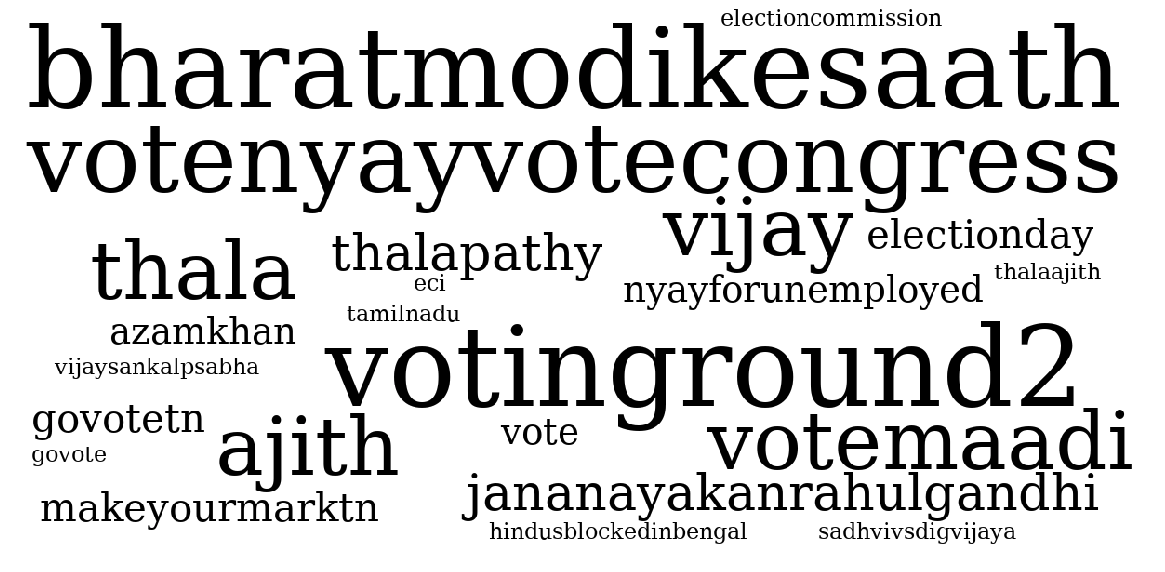}
    \caption{Phase 2} 
    \label{fig:top-hash-2} 
  \end{subfigure}
  \begin{subfigure}[b]{0.30\linewidth}
    \centering
    \includegraphics[width=\linewidth]{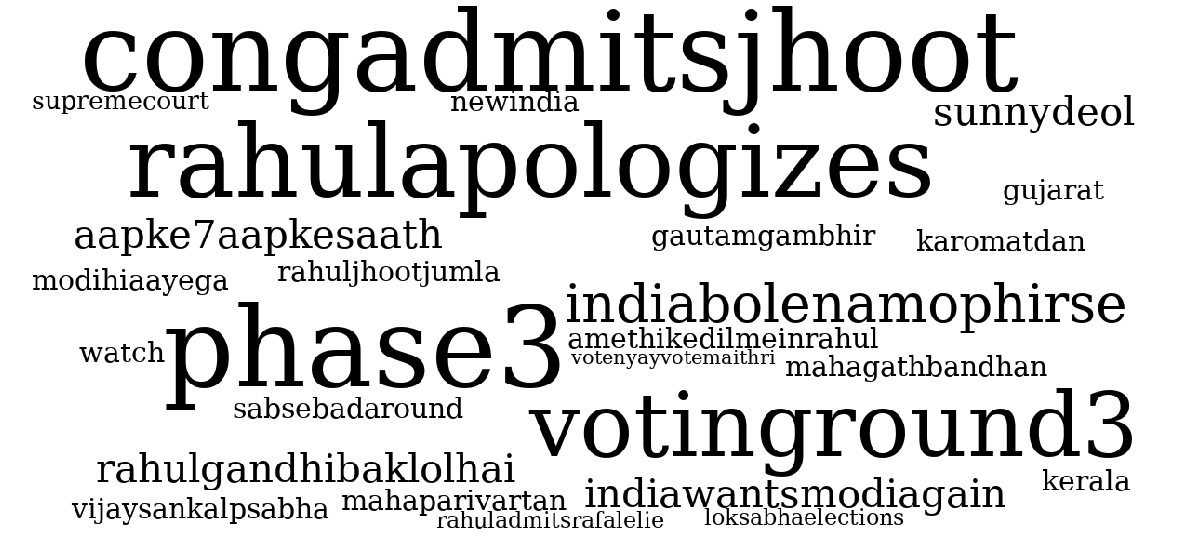}
    \caption{Phase 3} 
    \label{fig:top-hash-3} 
  \end{subfigure} 
  \begin{subfigure}[b]{0.30\linewidth}
    \centering
    \includegraphics[width=\linewidth]{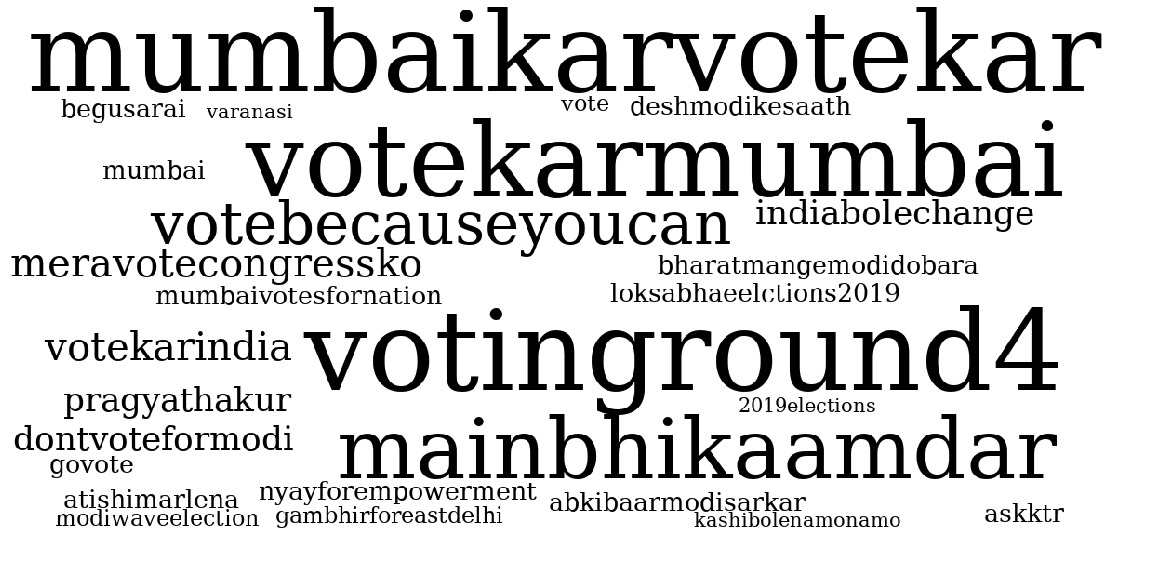}
    \caption{Phase 4} 
    \label{fig:top-hash-4} 
  \end{subfigure} 
  \begin{subfigure}[b]{0.30\linewidth}
    \centering
    \includegraphics[width=\linewidth]{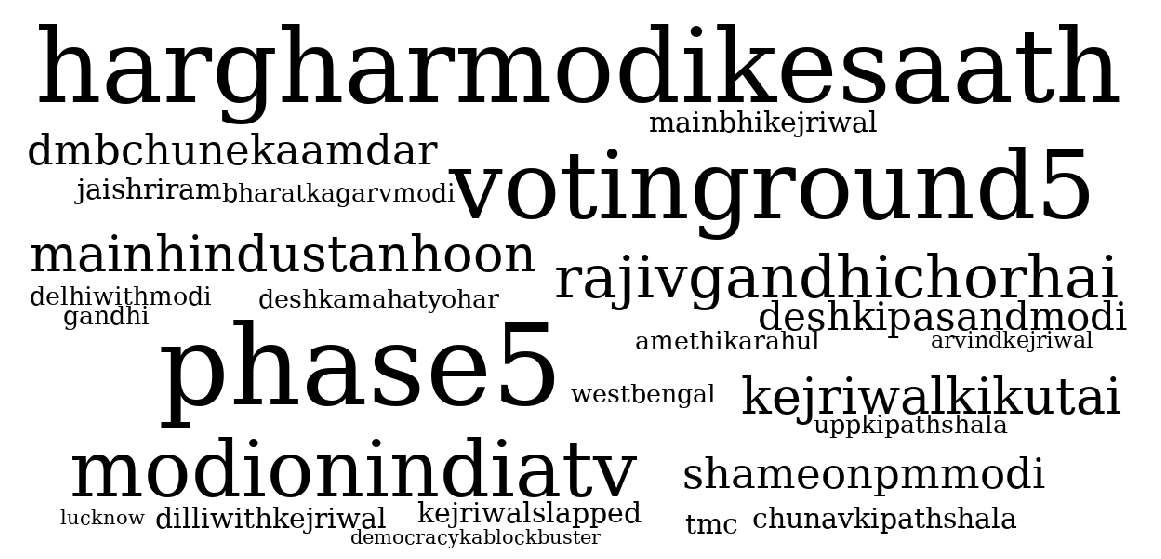}
    \caption{Phase 5} 
    \label{fig:top-hash-5} 
  \end{subfigure} 
  \begin{subfigure}[b]{0.30\linewidth}
    \centering
    \includegraphics[width=\linewidth]{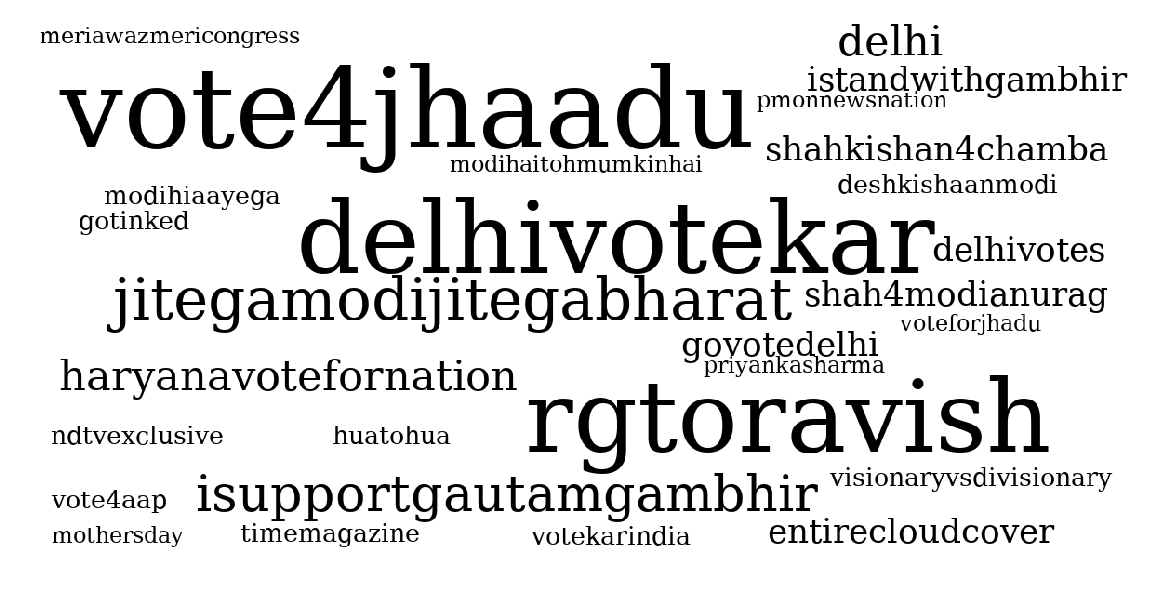}
    \caption{Phase 6} 
    \label{fig:top-hash-6} 
  \end{subfigure} 
  \begin{subfigure}[b]{0.30\linewidth}
    \centering
    \includegraphics[width=\linewidth]{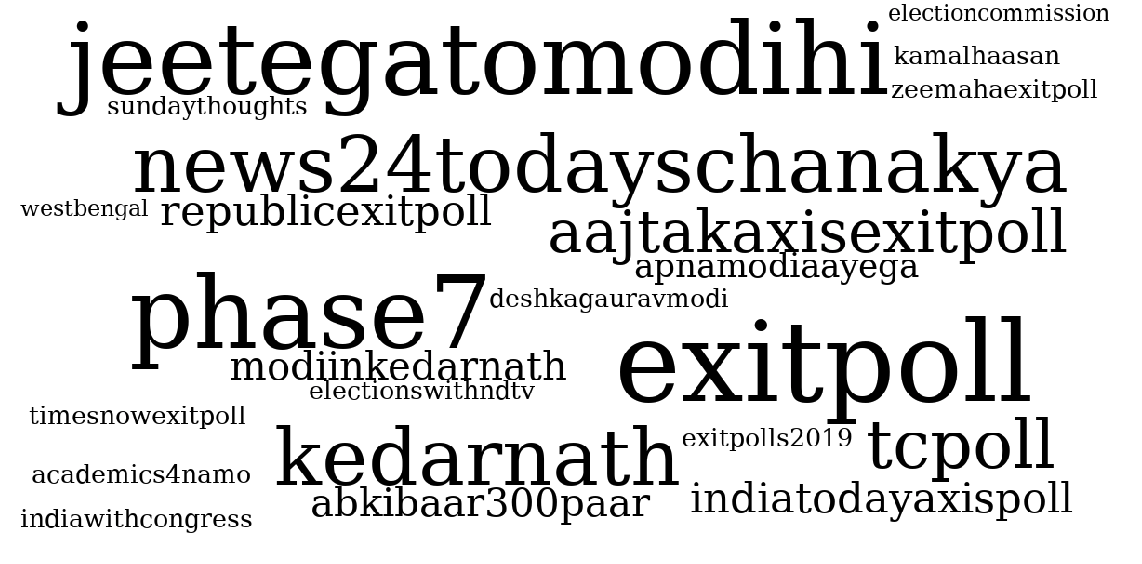}
    \caption{Phase 7} 
    \label{fig:top-hash-7} 
  \end{subfigure} 
  \caption{Top Hashtags over the course of elections and during each phases.}
  \label{fig:top-hash} 
\end{figure*}

For many experiments, we have divided the data into seven parts representing the seven phases. Each part has tweets corresponding to a particular phase duration. Figure \ref{fig:pvh} shows the distribution of number of hashtags in each tweet. Around 45\% tweets contain one or two hashtags. There are comparatively less number of tweets that contain at least 5 hashtags.

\section{Trends and Events}
\label{sec:trends}

In this section, we first portray some general trends using word clouds. We then utilize some top trends to fetch tweets that reveals the event unfolded on the ground around those trends.  

\subsection{Trends}
\label{sec:trs}
We draw a word cloud of the top 50 most occurring hashtags to show general trends in hashtags throughout elections. Figure \ref{fig:top-hash-all} shows curiosity for elections among Twitter users as they were using hashtags like \#votekar, \#voteforindia, \#loksabhaelections2019. 

Next, we divided the dataset into seven parts depicting each phase to observe general trends in the duration of each phase. Each part contains all the tweets posted within the duration of that particular phase. For example, tweets from Apr 11 to Apr 18 are segregated separately to draw the word cloud of hashtags in Phase 1. Figures \ref{fig:top-hash-1} - \ref{fig:top-hash-7} shows trends during phases 1-7, respectively. We removed the hashtags that are common between the overall trends (Figure \ref{fig:top-hash-all}) and phase-wise trends (Figure \ref{fig:top-hash-1} - Figure \ref{fig:top-hash-7}). The removal is necessary to differentiate overall trends from phase-wise trends. A behavior similar to the overall trends, is observed in the phase-wise plots as well. Hashtags are positively leaning more towards \#bjp as hashtags like \#isbaarnamophirse (translates to - Namo again this time), \#bharatmodiksath (translates to - India is with Modi), and \#hargharmodikesath (translates to - every home with Modi) are shared in large numbers. On the other hand, hashtags show mixed sentiments for other parties with hashtags like \#rahulapologizes, \#congadmitsjhoot (translates to - congress admits a lie), \#kejriwalslapped along with \#amethikedilmerahul (translates to - Rahul is in Amethi's heart), \#dilliwithkejriwal, \#indiawithcongress, and so on.

Qualitatively, the sentiment around trending hashtags is mixed. On the one hand, we have hashtags like \#modihaitomumkinhai, \#abkibaarmodisarkar, \#deshkipasandmodi, \#apnamodiaayega, while on the other hand, there are trending hashtags like \#saynotobjp, \#shameonpmmodi, \#indiawithcongress, etc.  


\begin{table*}[htbp]
    \centering
    \begin{tabular}{M{0.1\linewidth}|M{0.5\linewidth}|M{0.22\linewidth}}
    \hline
         \textbf{Date} & \textbf{Tweet} & \textbf{Trending Hashtag}  \\
    \hline
    
    April 11 & Is this the Election Commissions promise of free and fair election in West Bengal? You can ban a \#ModiBiopic but can not take action against goons. @WBPolice have you taken cognizance of this or are you hand in gloves with the ruling party in bengal. \#Vote4India \#VoteKar & \#modibiopic \\

    \hline
    Apr 21 & Rahul Gandhi can not speak or address masses but that apart he has not expressed regret for "chowkidar chor hai" but slogan but only for linking it to supreme court. Chowkidars try to understand. Bewaqoof ho kya. Padho samjho \#CongAdmitsJhoot \#RahulApologises & \#rahulapologizes \\
    \hline 
        May 5 & @RahulGandhi \#modi's comment on \#RajivGandhi is inappropriate n poor. That does not mean there were no scams during his reign. \#Bofors , landgrab, etc. But u don't speak ill of the dead. \#ModiAaneWalaHai \#RajivGandhiChorHai \#LokSabhaElections2019 \#NaMoAgain2019
 & \#RajivGandhiChorHai \\
    \hline 
        May 9 & \#IstandwithGautamGambhir @ArvindKejriwal bol do yeh bhi modi se Mila hua hai. Ab khud ke khode hue gadde Mei khud hi giroge @AtishiAAP \#DeshKiShaanModi \#LokSabhaElections2019 \#AAPtards https://t.co/gw8PVMy57p & \#istandwithgautamgambhir \\
    \hline 
    
        May 12 & Ready to vote today and woke up with the EntireCloudCover story of our very own Bal Narendra. This year I've stopped watching TheKapilSharmaShow and followed ModiInterviews. Sir kabhi disappoint nahi karte. & \#entirecloudcover \\
    \hline
    \end{tabular}
    \caption{Notable events cross-referenced in our dataset using trending hashtags. The tweet is fetched from the dataset and it contains the hashtag mentioned in the corresponding ``Trending Hashtag'' column.}
    \label{tab:events}
\end{table*}

\begin{table*}[!b]
    \centering
    \begin{tabular}{|M{0.4\linewidth}|M{0.4\linewidth}|}
    \hline
         \textbf{Topic} & \textbf{Topmost representative hashtags}  \\
    \hline
         Elections  & \#elections2019, \#loksabhaelections2019, \#vote, \#loksabhaelections \\
         \hline
         Promotions & \#voteforindia, \#votekar, \#vote4bjp, \#vote4modi \\
         \hline
         Modi Praise & \#phirekbaarmodisarkar, \#modioncemore, \#modihaitomumkinhai, \#namoagain \\
         \hline
         Dravida Munnetra Kazhagam (DMK, a political party in Tamil Nadu) & \#dmk, \#dmk4tn, \#votefordevelopment, \#dmkalliance \\
         \hline
         Congress & \#mahagathbandhan, \#rahulgandhi, \#wayanaad, \#pappu \\
         \hline
         BJP & \#chowkidar, \#savedemocracy, \#sankalphamaramodidubara, \#pmmodi \\
    \hline
    \end{tabular}
    \caption{Some topics found using Latent Dirichlet Allocation (LDA) results on hashtags. The topics are qualitatively assigned to each group of words. Note that the \#hashtags listed in the table are clustered in an unsupervised manner.}
    \label{tab:lda}
\end{table*}

\subsection{Events}
\label{sec:events}

We wanted to find the notable events that happened during the course of elections using just hashtags. For this, we used the trending hashtags generated from the previous section to cross-reference it with our dataset in order to track actual events that happened on the ground. 
We make a list of hashtags that are shared more than 8,000 times and randomly choose them to track down events. We search our dataset using the chosen hashtags to make a list of tweets containing each hashtag. Then we randomly choose a tweet that contains at least two sentences. Table \ref{tab:events} lists down a few of the notable tweets that are posted around each hashtag. The steps followed to get events are listed in Algorithm \ref{alg:events}.


\begin{algorithm}[htb!]
\caption{Pick Events}\label{alg:events}
\begin{algorithmic}[1]
\Procedure{$PickEvents$}{$hashtag$}\Comment{Top trending hashtag from Section \ref{sec:trends}}
\State $temp \gets []$  \Comment{Empty list}
\For{$tweet \in tweets $}
    \If{ $hashtag \; in \: tweet.text$ } 
        \State $temp \gets tweet$ \Comment{Add tweet to the list}
    \EndIf
\EndFor

\State $index \gets random(0, len(temp))$ \Comment{Choose random index for event}

\textbf{return} $temp[index]$\Comment{The chosen event for given hashtag}
\EndProcedure
\end{algorithmic}
\end{algorithm}

\begin{table*}[ht!]
    \centering
    \begin{tabular}{c|c|c|c}
    \hline
         \textbf{\#modi} & \textbf{\#raga} & \textbf{\#bjp} & \textbf{\#congress}   \\
    \hline
         \#modiji & \#pappu & \#mohanbhagwat & \#inc  \\
         \#primeminister & \#baildhari & \#amitshah & \#congressparty  \\
         \#bhakts & \#pappucongress & \#rss & \#cong  \\
         \#imrancampaignformodi & \#rafaeldeal & \#bjp4india & \#khangress  \\
         \#chowkidars & \#sorry & \#bjpnewstrack & \#aicc  \\
         \#surgicalstrike & \#chaukidaar & \#sambitpatra & \#tripletalaq \\
         \#surgicalstrike2 & \#chawkidar & \#roadshow & \#congressendgame \\
         \#airstrike & \#shameonyou & \#bhaiyyajijoshi & \#kpcc \\
         \#modiobsession & \#robertvadra & \#dhpoliticaltheatre & \#kirenrijiju \\
         \#modisarkar & \#pappu420 & \#modisarkar & \#hitler \\
    \hline
    \end{tabular}
    \caption{Top 10 most semantically similar hashtags to \#modi, \#raga, \#bjp, and \#congress during the course of elections2019.}
    \label{tab:semantic-close}
\end{table*}

\section{Topics and Semantics}
\label{sec:tns}

In this section, we used approaches to find the topics hidden among hashtags and see what hashtags are semantically similar to \#modi, \#raga, \#bjp, and \#congress\footnote{We choose these hashtags because \#bjp and \#congress are the two biggest parties with 435 and
420 candidates, participating in the elections. \#modi and \#raga are general terms used to mention leaders of the two parties, respectively.} during the course of elections. For this experiment, we filtered the tweets that have at least five hashtags. The number of hashtags in tweets vary from a minimum of 1 to a maximum of 7. After filtering, we were left with 212,935 tweets containing at least five hashtags. We treated each hashtag as a word and group of all hashtags in a tweet as a sentence. Further, we applied Latent Dirichlet allocation (LDA) to find latent topics and used skip-gram word embeddings to find semantically similar hashtags.

\subsection{Topic Modeling}
\label{sec:topics}

We wanted to experiment and find what topics are prominently related to Elections on Twitter. Topic modeling helped us analyze the overall user sentiment towards a party; it helped us understand the problems and opinions floating among Tweets. We used Latent Dirichlet Allocation (LDA) to find latent topics among the hashtags in an unsupervised manner. We chose LDA for the task because it calculates the document-topic and word-topic distributions using Dirichlet priors, which leads to better generalization ~\cite{Blei2003}. Dirichlet can be thought of as a distribution over a distribution. It helps us find an actual probability distribution based on a given type of distribution.

Performing LDA over the filtered dataset involved the following steps:

\subsubsection{Preprocessing}
To keep topic modeling based only on hashtags, we extracted the hashtags from each tweet. Each hashtag is treated as a token, i.e., a word. These tokens are then combined to form a sentence. The formed sentence acts like a document, which can be used to perform LDA. Usually, in prepossessing and feature selection, all the terms that occur a lesser number of times are removed along with the stop words. The remaining words are reduced to their lemmas to make topics more meaningful ~\cite{3817f31d07cb4731bc94befe3dc70721}. The way we create a document conforms with the preprocessing steps necessary to get good topics using LDA.

\subsubsection{Choosing no. of topics}
We need to choose the number of topics ($K$), which indicates how many latent topics the LDA model should find. There is no simple default value that can be chosen for this parameter. We aim to create a lower-dimensional representation for our data, which is big enough that we do not lose much relevant information. The problem with topic models is that they give no guarantee on the interpretability of their output. \citet{Roder:2015:EST:2684822.2685324} proposed a framework to calculate coherence measures to evaluate topic models. Therefore, to find the optimal number of topics to look for, we build many LDA models with a different number of topics ranging from $5$ to $40$ and calculated their coherence values. 

\begin{figure}[!b]
    \centering
    \includegraphics[width=0.8\linewidth]{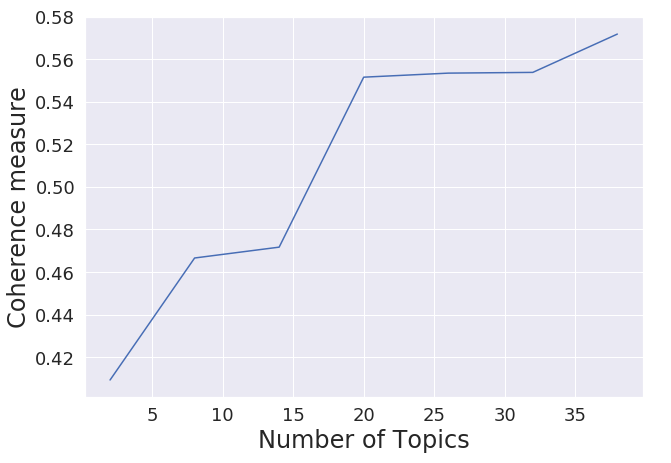}
    \caption{Coherence values of LDA models with different number of topics. The coherence value starts to converge at $K=20$. Therefore, we choose 20 as the number of topics to use with out LDA model.}
    \label{fig:coherence}
\end{figure}

Same keywords start to repeat in multiple topics if the value of $K$ is too large. Therefore, we choose the value of $K$ that lies at the starting point of the convergence of coherence values. As shown in Figure \ref{fig:coherence}, the optimal value of $K$ for our corpus is $20$.

\subsubsection{Topics}
After preprocessing the tweets and finding the optimal values for the number of topics, we run the LDA model on our dataset. Table \ref{tab:lda} shows some of these topics, along with the relevant keywords grouped using the LDA model. We manually choose a qualitative keyword to represent the grouped \#hashtags and assigned it as the topic. For example, hashtags like \#voteforindia, \#votekar (translates to - do vote), \#vote4bjp, and \#vote4modi are assigned \textbf{Promotions} as the topic. Table \ref{tab:lda} lists a few of them.

\subsection{Semantic Similarity}
\label{sec:semantic}

We wanted to find hashtags in our dataset that are semantically related to a query hashtag over the course of elections. We use the skip-gram model with negative sampling ~\cite{inproceedings13, Mikolov:2013:DRW:2999792.2999959} for the purpose. Skip-gram is used to maximize the similarity between the words which appear next to each other in the given corpus. It creates a continuous vector for each word in a manner that preserves a word's context. 

For this experiment, we extract all the hashtags present in a tweet. Each hashtag is treated as a word. Combining all hashtags from each tweet forms a sentence. Then we perform skip-gram analysis over all the sentences. ~\citet{hamilton-etal-2016-diachronic} empirically shows that a context vector of dimension 300 gives the best results. Therefore, based on previous works, we set the dimension of each hashtag's vector to 300. 

We used cosine distance as the similarity metric and chose four hashtags: \#modi, \#raga, \#bjp, \#congress to perform a qualitative analysis to find hashtags that are most semantically similar to these four. We chose these hashtags because the Bhartiya Janata Party (BJP) and the Indian National Congress (or just congress) are two biggest parties with 435 and 420 candidates ~\cite{manish} participating in the elections, respectively. The hashtag \#modi refers to Narendra Modi (BJP leader), and \#raga refers to Rahul Gandhi (Congress leader). Table \ref{tab:semantic-close} shows the results for \#modi, \#raga, \#bjp, \#congress. As we can see, the hashtags found from the experiment indeed are quite similar, e.g., \#raga is similar to \#pappu ~\cite{prabhash}, \#robertvadra ~\cite{indiatoday} and \#bjp is similar \#modisarkar, \#sambitpatra (BJP candidate), and \#rss ~\cite{admin}.

\subsection{Who's winning the Social Media Battle?}
\label{sec:battle}
We wanted to correlate the popularity of a candidate on Twitter to the election outcomes (win/lose) based on hashtags. To study this, we first searched for ``vs'' and ``versus'' in the hashtags present in our dataset to find the ones that represent two candidates competing against each other in elections. Our exhaustive search ended up getting us 8 candidates, but we looked at the top three most prominent ones like \#smritiiranivsrahulgandhi, \#gautamgambhirvsatishimarlena, and \#sadhvipragyavsdigvijay.

\begin{figure}[!b]
    \centering
    \includegraphics[width=\linewidth]{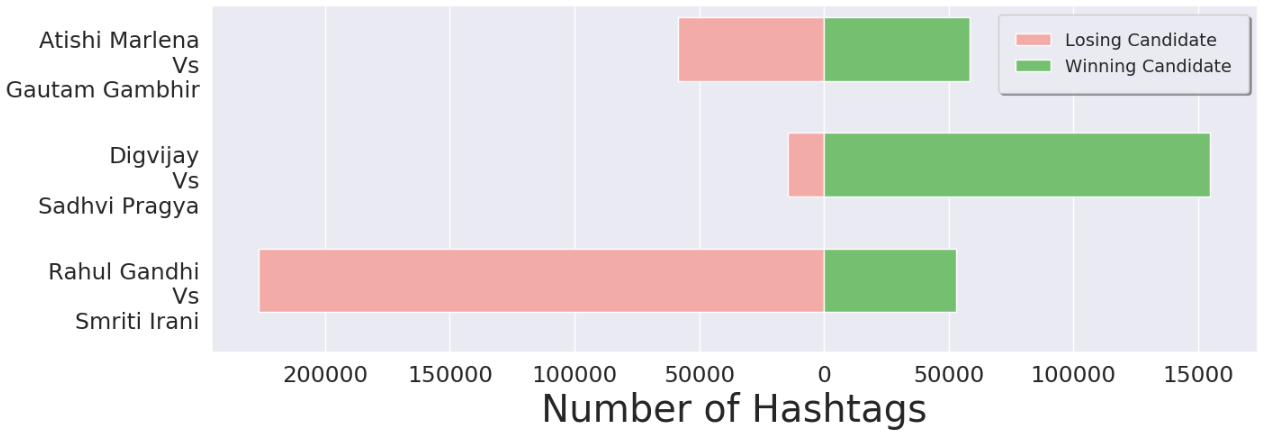}
    \caption{The popularity of a candidate on Twitter. The winning candidate is shown in green and the loosing candidate in red. The popularity value of Rahul Gandhi, Atishi Marlena, and Digvijay is on the left side, shown in red. Similarly, popularity value of Smriti Irani, Gautam Gambhir, and Sadhvi Pragya is shown in green on the right.}
    \label{fig:popularity}
\end{figure}

\begin{figure*}[!b]
    \centering
    \includegraphics[width=0.6\linewidth]{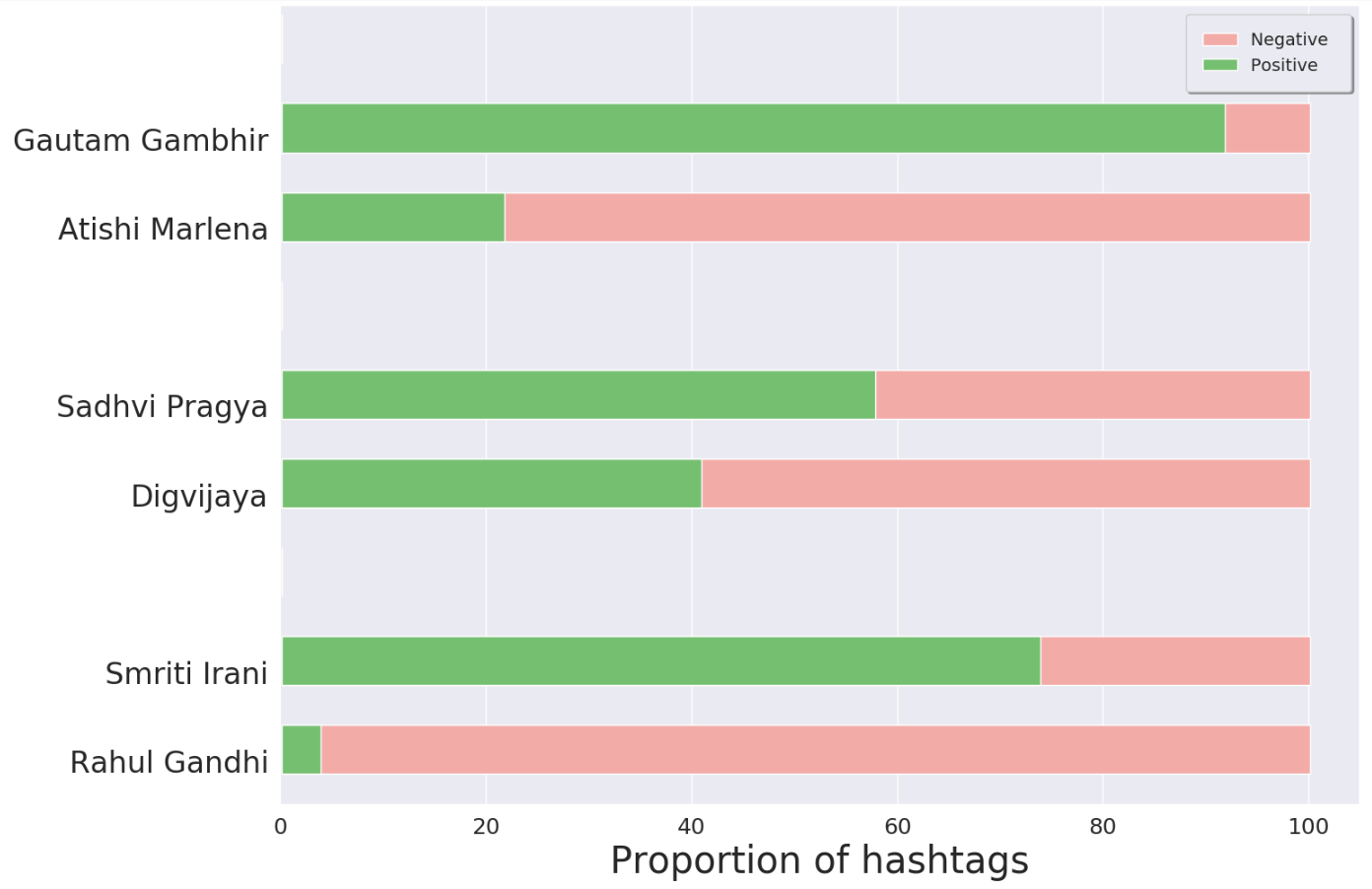}
    \caption{The percentage of negative and positive hashtags for key battles. The proportion of positive hashtags for each winning candidate is greater than the negative one, rendering them a higher influence score. }
    \label{fig:posneg}
\end{figure*}

Using the skip-gram analysis, as discussed in Section \ref{sec:semantic}, we first find the top 10 most semantically similar hashtags using them (hashtags with vs) as query hashtag. For example, \#smritiiranivsrahulgandhi fetched similar hashtags for both candidates. Some of them are \#rahulgandhiwayanad, \#amethi, \#rahulgandhiinterview, etc\footnote{Rahul Gandhi participated in elections from Amethi and Wayanad constituency.}, and \#amethi, \#amethikibetismriti, \#amethiwithsmritidi, etc.\footnote{Smriti Irani participated in elections from Amethi constituency.} To proceed, we removed the hashtags common for both candidates (\#amethi in the given example) and then calculated a sum over the occurrences of each hashtag.


If $hashtags[1,2,..,n]$ represents a list of semantically similar hashtags, we find $C_{hashtag[i]}$, the number of times a hashtag $i$ appeared and define popularity as the sum of all $C_{hashtag[i]}$'s. The popularity is given as:

\[ popularity \; = \; \sum_{i=1}^{n} C_{hashtag[i]}  \]

Figure \ref{fig:popularity} shows that the popularity on social media is not directly proportional to winning of a candidate. The popularity value of Rahul Gandhi, Atishi Marlena, and Digvijay is on the left side, shown in red. Similarly, popularity value of Smriti Irani, Gautam Gambhir, and Sadhvi Pragya is shown in green on the right. We have Rahul Gandhi with hashtags shared more number of times, but he lost in Amethi against Smriti Irani \cite{indiato}. The trend among the other battles is contrasting. We have Gautam Gambhir and Atishi Marlena with almost similar popularity, and Digvijay and Sadhvi Pragya showing a positive correlation between popularity and election outcome. 

Among the semantically similar hashtags, we have three categories of hashtags viz. positive, negative and neutral. We are not able to see a clear trend using the popularity scores of candidates. Therefore, we hypothesized that adding sentiment feature to popularity score might improve the results. 

To add sentiment information, we manually annotated the semantically similar hashtags as positive, negative or neutral. Then we find the number of occurrences, $C_{hashtag[i]}^+$ for all positive and $C_{hashtag[i]}^-$ for all negative hashtags for all candidates to calculate the influence, given as:  

\[ influence \; = \; \sum_{i=1}^{x} C_{hashtag[i]}^+ -  \sum_{i=1}^{y} C_{hashtag[i]}^-  \]

The influence value is a better metric than popularity as it also takes qualitative measures into account. Instead of considering only the count of similar hashtags, influence measure takes into account the sentiment of hashtags as well.

Figure \ref{fig:posneg} validates our hypothesis as we can see, Rahul Gandhi has more negative hashtags than the positives ones while on the other hand, Smriti Irani has more positive hashtags than the negative ones. In other words, the influence score of Rahul Gandhi is negative, i.e., smaller than Smriti Irani. The same pattern is true for the other two candidates as well. 






\section{Related Work}
\label{sec:literature}

Hashtags allowed researchers to study the behaviour and patterns of modern society on several online social media like Twitter and Instagram.   
There are studies done using just hashtags across several OSNs like Twitter, Instagram, and so on. \citet{Ferragina2015} focused on problems related to the meaning of hashtags. They make a Hashtag-Entity graph to model co-occurrences and semantic relatedness among hashtags and entities on Twitter to perform two natural language tasks - hashtag relatedness and hashtag classification. Moreover, their approach outperforms the state-of-the-art by a vast margin. \citet{Tsur2012} presented an approach that used linear regression to predict the spread of an idea in a given time frame. They evaluated their algorithm on Twitter hashtags extracted from 400+ million tweets.  ~\citet{Zhang:2019:LOT:3308558.3313480} did a comprehensive general analysis of hashtags on Instagram, which is a very different OSN from Twitter. The author performs extensive experiments using just hashtags to understand the temporal and spatial patterns, semantic displacement, and infer social relations among users. 

In the works that study a specific type of non-political hashtag, ~\citet{journals/corr/MejovaAH16} collected data based on the hashtag \#foodporn on Instagram to answer the question - whether \#foodporn promotes unhealthy food habits. They show nationwide trends of behavior using demographic analysis. Based on data from Twitter, \citet{Davidov:2010:ESL:1944566.1944594} creates a supervised framework for sentiment classification. The authors used 50 Twitter tags and 15 smileys as class labels and focused on creating rich features from patterns and punctuations. Their framework did not beat the human evaluation, but they provide a good enough starting point for the problem.  

In studies that focus on political hashtags, \citet{Small2011} used \#cdnpoli to analyze content on Twitter and perform: i) a categorization on types of users who use political hashtags, ii) investigate the nature of tweets tagged using \#cdnpoli, and iii) to what extent Twitter plays a role on political conversations. The author uses the content in the tweets for a lot of these experiments. Similarly, ~\citet{articlemunde} studied a socio-political movement using Twitter involving racial discrimination and police violence, known as Black Lives Matter (BLM). The authors collected data around the hashtag \#blacklivesmatter from Aug 2014 to May 2015, dividing it into four phases. They studied the temporal characteristics of participation on social media in the BLM movement, geographic spread of engagement and linguistic attributes, and how social media attributes and language relate to events unfolded on the ground. Besides performing a broader set of experiments on political hashtags, we use only hashtags, unlike ~\citet{Small2011} and ~\citet{articlemunde} who takes into account other attributes of a Tweet as well. 

\section{Discussion}
\label{sec:conclude}
In this paper, we perform a large-scale empirical study of political hashtags from Indian context on Twitter. We wanted to analyze what all patterns we can observe using only the hashtags from data collected during Loksabha elections 2019 in India. We collected 24.9 million hashtags from 8.22 million tweets for this study. We divided the dataset into seven phases of elections to perform our analysis.

We first show the trending hashtags from each phase. We take one of the most trending hashtags in a phase and cross-reference to our dataset to find the actual event that unfolded on the ground. Even after making a random selection in our algorithm, we find good quality tweets from the data. Tweets clearly explained the event; therefore, testifying about the excellent quality of our data. Moreover, we use Latent Dirichlet Allocation (LDA) to assign qualitative topics among hashtags. The topics help us qualitatively cluster the hashtags. We then use the skip-gram word embeddings to find that \#pappu is semantically similar to \#raga, and \#modi is accredited with \#airstrike, \#surgicalstrike.  We also compared the popularity of some major candidates participating in the elections and contrasted their election outcome using hashtags that are semantically similar to their names. We proposed an influence score, based only on hashtags, that can predict the election outcomes using just the hashtags. 

\bibliographystyle{ACM-Reference-Format}
\bibliography{sample-authordraft}










\end{document}